\documentclass[12pt]{article}

\usepackage{amsmath,amssymb,bm,natbib,fullpage}

\begin{document}

\title{Discussion of Peters, B\"{u}hlmann and Meinshausen}
\author{Chris. J. Oates$^{1,2}$, Jessica Kasza$^{3,4}$ and Sach Mukherjee$^5$\\
$^1$University of Technology Sydney\\
$^2$ARC Centre of Excellence for Mathematical and Statistical Frontiers\\
$^3$Monash University\\
$^4$Victorian Centre for Biostatistics\\
$^5$German Center for Neurodegenerative Diseases, Bonn.}
\maketitle

\begin{abstract}
Contribution to the discussion of the paper ``Causal inference using invariant prediction: identification and confidence intervals'' by Peters, B\"{u}hlmann and Meinshausen, to appear in the Journal of the Royal Statistical Society, Series B.
\end{abstract}

The information theorist is taught to list invariances and then derive models that exhibit those invariances.
We warmly congratulate the authors on their insight and creativity in bringing ideas of invariance to bear on causal inference and, in so doing, furthering our understanding of the connection between causality and prediction, two areas that are often treated as  separate (if not opposed!) but that are in fact closely related. 

An attractive feature of the proposed method is that it allows integration of multiple data sources in a natural way, even when the precise nature or target of perturbations to the data are unknown. On the other hand, in some cases  a softer approach that allows for  variation in causal structure across data subsets might be appropriate \citep[see e.g][]{Oates3}.

Returning to invariance, we wonder how far we can push the information theorist.
For instance, consider estimation of the causal effect $\theta_{ij}$ of one variable $X_i$ on another $X_j$.
A correct causal graph $G$ can be interrogated to produce a minimal sufficient set $S$ of variables to adjust for when estimating  $\theta_{ij}$ \citep{Pearl}. 
These variables $S$ can then be included in a propensity score model, leading ultimately to an estimate $\hat{\theta}_{ij}(S)$ for $\theta_{ij}$.
Often such minimal sufficient adjustment sets are not unique, in which case any other minimal sufficient adjustment set $S'$ will do (in the sense of allowing consistent estimation).
Then, we would expect, for large sample sizes, 
$\hat{\theta}_{ij}(S) \approx \hat{\theta}_{ij}(S').$
    On the other hand, there seems no particular reason to expect that these two estimates would coincide if the graph $G$ were incorrect.
This seems to suggest another invariance that could be exploited for causal discovery.
Potentially, other invariances could play a role.
This may in future lead to having to ask which invariances are most useful in practice.

There has long been (in our view justifiable) empirical skepticism towards {\it de novo} causal discovery, a skepticism expressed by, among others, David Freedman and Paul Humphreys 
in a paper titled ``Are there algorithms that discover causal structure?" \citep[][]{Freedman}. The issue is that it is difficult to empirically validate causal discovery on 
a given problem using data at hand, leaving the analyst unsure as to whether or not the output of a given procedure should be trusted. 
This goes a bit further than familiar  issues of statistical uncertainty, since the underlying concern is of a potentially profound mismatch between critical assumptions and the real data-generating system. The authors' insightful discussion of model mis-specification is therefore welcome and the conservative behaviour of their procedure very appealing. We note also that background scientific knowledge may itself be mis-specified but that in some circumstances it may be possible to effect ``repairs" on the relevant causal structures \citep{Oates}. 
We see it as a very positive development that  empirical validation of causal discovery is becoming more common \citep[see e.g.][]{Hill}.  In the near future, empirical work, not least in biology, ought to give us a better sense of the practical efficacy of causal discovery. A concrete  answer to Freedman and Humphreys' question may then start to come into reach.

\end{document}